\documentclass[twocolumn,showpacs,preprintnumbers,prl,nofootinbib]{revtex4-1}
\usepackage{graphicx,epsfig,amsmath,amssymb,bm}
\usepackage{color}
\makeatletter
\DeclareRobustCommand*{\bfseries}{%
  \not@math@alphabet\bfseries\mathbf
  \fontseries\bfdefault\selectfont
  \boldmath
}
\makeatother

\begin{document}

\preprint{MZ-TH/11-50} 

\title{Bounds on Warped Extra Dimensions from a Standard Model-like Higgs Boson}

\author{Florian Goertz$^a$, Ulrich Haisch$^b$ and Matthias Neubert$^c$} 

\affiliation{${}^a$\,Institute for Theoretical Physics, ETH Zurich, 8093 Zurich, Switzerland\\
${}^b$\,Rudolf Peierls Centre for Theoretical Physics, University of Oxford, 
OX1 3PN Oxford, United Kingdom\\
${}^c$\,Institut f\"ur Physik, Johannes Gutenberg-Universit\"at,
D-55099 Mainz, Germany}

\date{December 21, 2011}

\begin{abstract}
We point out that the discovery of a light Higgs boson in the $\gamma\gamma$, $ZZ$ and $WW$ decay channels at the LHC, with cross sections not far from the predictions of the Standard Model, would have important implications for the parameters of warped extra-dimension models. Due to loop effects of Kaluza-Klein particles, these models predict a significant reduction of the Higgs production cross section via gluon-gluon fusion, combined with an enhancement of the ratio $\mbox{Br}(h\to\gamma\gamma)/\mbox{Br}(h\to ZZ)$. LHC measurements of these decays will probe Kaluza-Klein masses up to the 10\,TeV range, exceeding by far the reach for direct production.
\end{abstract}

\maketitle

\subsubsection{Introduction}

The announcement of the first evidence for a Higgs-boson signal in the search channels $h\to\gamma\gamma$, $h\to ZZ^*\to ll\bar l\bar l$, and $h\to WW^*\to l\bar l\nu\bar\nu$ by the ATLAS and CMS collaborations at CERN \cite{CERNtalks} defines a turning point in the history of elementary-particle physics. Even though the significance of the signal is not yet sufficient to claim a discovery, it now seems likely that the existence of a Higgs boson will be confirmed as early as in 2012. This discovery will mark the birth of the hierarchy problem \cite{Gildener:1976ai,Weinberg:1978ym}. In the years to come the focus will be on probing the properties of the Higgs boson in as much detail as possible, so as to see whether there are any deviations from the predictions of the Standard Model (SM).

Immediately following the announcement of the LHC results, several authors have commented on their implications for the SM \cite{Wetterich:2011ct,Englert:2011er,EliasMiro:2011ex,Xing:2011ie}, its supersymmetric extensions \cite{Baer:2011eu,Feng:2011ew,Li:2011ez,Heinemeyer:2011fa,Arbey:2011fb,Draper:2011hd,Carena:2011mw,Ellwanger:2011sv,Buchmueller:2011tb,Akula:2011up,Kadastik:2011ur,Strege:2011pk,Cao:2011sn}, and other new-physics models \cite{Guo:2011jm,Ferreira:2011kt,Burdman:2011ki,Cheung:2011nv}. In particular, it was pointed out that a Higgs mass near 126\,GeV is such that the SM can be extrapolated up to very high mass scales without running into trouble with triviality or vacuum instability bounds. If one is willing to accept a metastable vacuum with lifetime exceeding the age of the Universe, then the SM may even be extrapolated all the way to the Planck scale \cite{EliasMiro:2011ex}. The hierarchy problem may then be solved by some non-trivial dynamics, e.g.\ a fixed-point of some renormalization-group flow, at very high energy scales \cite{Wetterich:2011ct}. In such a situation the physics responsible for stabilizing the Higgs-boson mass may lie out of the reach of terrestrial collider experiments.

Despite this possibility, extensions of the SM invoking new particles at the TeV scale remain, of course, the more attractive way to address the hierarchy problem. Besides supersymmetry, models with extra dimensions provide an interesting class of scenarios, in which the hierarchy between the Planck and electroweak scales is explained in terms of geometry. Randall-Sundrum (RS) models \cite{Randall:1999ee}, in which the extra dimension is an anti-de Sitter (AdS) space, are particularly compelling in this regard, since they can address both the hierarchy and the flavor problem -- the question why the fermion masses and mixing angles exhibit large hierarchies -- in terms of the same geometrical mechanism \cite{Grossman:1999ra,Gherghetta:2000qt,Huber:2000ie,Huber:2003tu,Agashe:2004cp}.

In this letter, we point out that a Higgs-boson discovery at the LHC will have important implications for RS models with a Higgs sector localized on (or near) the infra-red brane, while matter and gauge fields are allowed to propagate in the bulk.  We have emphasized in \cite{Casagrande:2010si} that in the context of these models the Higgs production cross section via gluon-gluon fusion and the Higgs branching fractions provide superb tools to probe for indirect signals of new physics, even if the mass scale of the new particles lies outside the reach for direct production at the LHC. These models predict significant modifications of the loop-induced Higgs couplings to gluons or photons even for Kaluza-Klein (KK) masses in the range of several TeV, due to loop diagrams with virtual exchanges of KK fermions or $W$ bosons. We found that the Higgs-boson production cross section at the LHC can be significantly suppressed by these effects, while the branching fraction for Higgs decay into two photons is enhanced. On the other hand, the branching fractions for decays into $ZZ$, $WW$, and  $b\bar b$ remain largely unaffected. They are dominated by the tree-level couplings of the Higgs boson to fermions and gauge bosons, which remain almost SM-like. As a result, the cross section times branching ratios for $pp\to h\to\gamma\gamma$ and $pp\to h\to ZZ, WW$ are all suppressed, but the suppression is more pronounced in the latter case. Tentatively, this seems to be in accordance with the preliminary findings reported by ATLAS and CMS, which in particular indicate that any deviation from the SM predictions must be of moderate size.

For completeness, we note that different predictions for Higgs phenomenology in RS models were obtained in \cite{Azatov:2010pf}, where the authors find an enhancement of the effective $ggh$ coupling accompanied by a suppression of the $\gamma\gamma h$ vertex. We will explain elsewhere why we believe that these are not the correct results for RS models addressing the hierarchy problem, which must be considered as effective theories with a natural ultra-violet cutoff \cite{inprep}.

In order to interpret the LHC data in the context of a specific RS model, it is desirable to have available approximate analytical expressions connecting the Higgs-boson production and decay rates with fundamental parameters of the underlying five-dimensional (5D) theory. Such formulae will be derived in \cite{inprep}. Here we present the most important results of this analysis, sometimes making approximations which help to better illustrate the structure of the results. The most important model parameters entering our analysis are the KK scale $M_{\rm KK}=k\epsilon$ (with $k$ the AdS curvature and $\epsilon=\Lambda_{\rm IR}/\Lambda_{\rm UV}\approx 10^{-16}$) and the 5D Yukawa matrices $\bm{Y}_u$ and $\bm{Y}_d$ in the up and down sectors (these matrices were called $\tilde{\bm{Y}}_{u,d}$ in \cite{Casagrande:2010si}). We recall that the flavor hierarchies observed in the mass spectrum and mixing angles of the SM quarks can be explained naturally in terms of anarchic 5D Yukawa couplings, i.e.\ complex-valued matrices with random elements, and wave-function overlap integrals. The KK scale determines the masses of the low-lying KK excitations. For instance, the mass of the lightest KK gluon is approximately 2.45\,$M_{\rm KK}$. We also define the ``volume'' $L=\ln(1/\epsilon)\approx 37$ of the extra dimension. For simplicity we will study the minimal RS model with matter and gauge fields propagating in the bulk, making use of many results derived in \cite{Casagrande:2008hr}. We will later comment on an extended model with a custodial protection of electroweak precision observables \cite{Agashe:2003zs, Agashe:2006at}, which was the basis of our previous study of Higgs physics in a warped extra dimension \cite{Casagrande:2010si}.

\subsubsection{Higgs-boson production}

The Higgs-boson production cross section in gluon-gluon fusion at the LHC can be written as
\begin{equation}
    R_h = \frac{\sigma(gg\to h)_{\rm RS}}{\sigma(gg\to h)_{\rm SM}}
    = \frac{\kappa_g^2}{\kappa_v^2} \,,
\end{equation}
where 
\begin{equation} \label{eq:kappav}
   \kappa_v = \frac{v_{\rm RS}}{v_{\rm SM}}
   \approx 1 + \frac{m_W^2}{4M_{\rm KK}^2}\,(L-1)
\end{equation}
is a small correction factor accounting for the difference of the Higgs vacuum expectation value $v_{\rm RS}$ in the minimal RS model from its value in the SM \cite{Bouchart:2009vq}. For a KK scale of 2\,TeV, we find $\kappa_v\approx 1.014$. Furthermore,
\begin{equation}\label{kappag}
\begin{aligned}
   \kappa_g &\approx \mbox{Re}(\kappa_t) - \sum_{q=u,d}\,\mbox{Tr}\,f(\bm{X}_q) \,, \\
   f(x) &= \frac{x(1-x^2)}{1+x^2}\,\tanh^{-1}x
    = x^2 - \frac{5x^4}{3} \pm \dots
\end{aligned}
\end{equation}
represents the modifications to the loop-induced $ggh$ coupling \cite{inprep}. We neglect the contribution of the bottom quark, which is strongly suppressed, and replace the top-quark loop function by its asymptotic value 1, which is an excellent approximation for a light Higgs boson. The sum accounts for the one-loop contributions of KK quarks. Remarkably, the infinite sum over these modes converges and its value depends only on the KK scale $M_{\rm KK}$ and the fundamental Yukawa matrices of the 5D theory, via the variable
\begin{equation}
   \bm{X}_q = \frac{v_{\rm RS}}{\sqrt 2 M_{\rm KK}}\,\sqrt{\bm{Y}_q\,\bm{Y}_q^\dagger} \,.
\end{equation}
The hermitian matrices $\bm{Y}_q\,\bm{Y}_q^\dagger$ can be diagonalized by means of a unitary transformation. After taking the trace in generation space, we obtain
\begin{equation}
   \mbox{Tr}\,f(\bm{X}_q) 
   = \sum_i\,f\bigg(\,\frac{v_{\rm RS}\,|y_q^{(i)}|}{\sqrt2 M_{\rm KK}}\bigg) \,,
\end{equation}
where $y_q^{(i)}$ are the eigenvalues of the Yukawa matrices. Since the 5D Yukawa matrices are assumed to be non-hierarchical with ${\cal O}(1)$ complex elements, it follows that the result is proportional to the number of fermion species running in the loop. In other words, the KK towers of all quarks give comparable contributions to the effective $ggh$ vertex, irrespective of the mass of the corresponding SM quark.

In RS models the Higgs-boson coupling to top quarks is modified with respect to its SM value, since the top quark mixes with its KK excitations. This effect is accounted for by the parameter $\kappa_t$ in (\ref{kappag}), whose real part is always less than 1. We obtain 
\begin{equation}\label{kappat}
   \kappa_t\approx 1 - \frac{v_{\rm RS}^2}{3M_{\rm KK}^2}\,
    \frac{\big(\bm{Y}_u\bm{Y}_u^\dagger\bm{Y}_u\big)_{33}}{\big(\bm{Y}_u\big)_{33}} 
    - \big( \bm{\delta}_U \big)_{33} - \big( \bm{\delta}_u \big)_{33} \,.
\end{equation}
Expressions for the matrices $\bm{\delta}_{U,u}$ can be found in \cite{Casagrande:2008hr}. 

\subsubsection{Higgs-boson decays}

We focus on the decays of the Higgs boson into the di-boson final states $f = ZZ$, $WW$, $\gamma\gamma$, which for a light Higgs boson will be measured with good precision at the LHC. In the minimal RS model the corrections to the $ZZ$ and $WW$ final states are identical up to a factor $m_Z^2/m_W^2$. We define corrections factors ($V=Z,W$)
\begin{equation}
\begin{aligned}
   \frac{\Gamma(h\to VV)_{\rm RS}}{\Gamma(h\to VV)_{\rm SM}}
   &= \kappa_v^2\,\kappa_V^2 \,, \\
   \frac{\Gamma(h\to\gamma\gamma)_{\rm RS}}{\Gamma(h\to\gamma\gamma)_{\rm SM}}
   &= \frac{\kappa_\gamma^2}{\kappa_v^2} \,,
\end{aligned}
\end{equation}
where to a good approximation \cite{Casagrande:2010si}
\begin{equation} \label{eq:kappaV}
   \kappa_V \approx 1 - \frac{m_V^2}{M_{\rm KK}^2}\,(L-1) \,,
\end{equation}
which yields $\kappa_Z\approx 0.925$ and $\kappa_W\approx 0.942$ for $M_{\rm KK}=2$\,TeV. The correction for the di-photon final state has a richer structure. We obtain
\begin{eqnarray}\label{eq:kappaga}
\begin{split}
   \kappa_\gamma 
   &\approx \frac{1}{A_W-\frac{4N_c}{9}}\,\Bigg[
    \kappa_W A_W + \frac{21}{8}\,\frac{m_W^2}{M_{\rm KK}^2}\,(L-1) \\[1mm]
   &\quad\mbox{}- N_c \left( \frac49\,\kappa_g  + \frac13\,\mbox{Tr}\,f(\bm{X}_d)
    - \frac{1}{N_c}\,\mbox{Tr}\,f(\bm{X}_l) \right) \! \Bigg] \,, \hspace{4mm}
\end{split}
\end{eqnarray}
where $A_W\approx 6.27$ is the absolute value of the $W$-boson loop function for $m_h=126$\,GeV, and $N_c=3$ is the number of colors. The first two terms inside the bracket correspond to the contributions of the $W$ boson and its KK tower, while the last term accounts for the contributions of the top quark as well as the KK quarks and charged leptons. Since in anarchic RS models the 5D Yukawa matrices are random matrices with elements of the same magnitude, the last two terms in parenthesis tend to cancel each other, so that we obtain numerically
\begin{equation}
   \kappa_\gamma\approx 1 + 0.27\,(1-\kappa_g) - 0.74\,(1-\kappa_W) \,.
\end{equation}
This result is interesting, because it shows that for certain choices of parameters the $h\to\gamma\gamma$ decay rate can be enhanced relative to its SM value.

When converting these results into branching fractions, one must take into account that the total decay rate $\Gamma(h)$ of a light Higgs boson is dominated by its decay into bottom quarks. For a SM Higgs with mass $m_h=126$\,GeV, one finds $\mbox{Br}(h\to b\bar b)\approx 0.59$, $\mbox{Br}(h\to WW)\approx 0.23$, $\mbox{Br}(h\to gg)\approx 0.07$, and $\mbox{Br}(h\to ZZ)\approx 0.03$. Hence 
\begin{equation} 
\begin{aligned}
   R_\Gamma & = \frac{\Gamma(h)_{\rm RS}}{\Gamma(h)_{\rm SM}} \\ 
   &\approx 0.59\,\big[\mbox{Re}(\kappa_b)\big]^2 + 0.26\,\kappa_W^2 
    + 0.07\,\kappa_g^2 + 0.09 \,.
\end{aligned}
\end{equation}
The correction factor $\kappa_b$ is approximately given by \cite{Casagrande:2010si}
\begin{equation}\label{kappab}
   \kappa_b\approx 1 - \frac{v_{\rm RS}^2}{3M_{\rm KK}^2}\,
   \frac{\big(\bm{Y}_d\bm{Y}_d^\dagger\bm{Y}_d\big)_{33}}{ \big(\bm{Y}_d\big)_{33} } \,.
\end{equation}

We are now in a position to work out the products of the production cross section times branching ratios for the various decay channels. These are the key observables that will be affected by new-physics contributions in RS scenarios. Defining the ratios
\begin{equation}
   R_f = \frac{\big[\sigma(pp\to h)\,\mbox{Br}(h\to f)\big]_{\rm RS}}%
              {\big[\sigma(pp\to h)\,\mbox{Br}(h\to f)\big]_{\rm SM}} \,,
\end{equation}
we obtain 
\begin{equation}\label{Rgg}
   R_{VV} = \frac{\kappa_g^2\,\kappa_V^2}{R_\Gamma} \,, \qquad 
   R_{\gamma\gamma} 
   = \frac{\kappa_g^2\,\kappa_\gamma^2}{\kappa_v^4\,R_\Gamma} \,,
\end{equation}
from which it follows that $R_{\gamma\gamma}/R_{VV}=\kappa_\gamma^2/(\kappa_v^4\,\kappa_V^2)$. 

\subsubsection{Phenomenological implications}

The most significant corrections are likely to arise from the terms involving the Yukawa matrices, i.e.\ $\kappa_g$ in (\ref{kappag}), as well as $\kappa_t$ and $\kappa_b$ in (\ref{kappat}) and (\ref{kappab}). By randomly generating a large set of 5D Yukawa matrices, whose individual entries are restricted to have absolute values below $y_{\rm max}$ and are required to reproduce the masses and mixing angles observed in the quark sector, we find that on average 
\begin{equation}\label{eq:mRS}
\begin{aligned}
   \mbox{Tr}\,\bm{X}_q^2 &\approx \frac{v_{\rm RS}^2}{2M_{\rm KK}^2}\,3.1\,y_{\rm max}^2 \,, \\
   \kappa_t &\approx 1 - \frac{v_{\rm RS}^2}{2M_{\rm KK}^2}\,1.4\,y_{\rm max}^2 \,, \\
   \kappa_b &\approx 1 - \frac{v_{\rm RS}^2}{2M_{\rm KK}^2}\,1.0\,y_{\rm max}^2 \,.
\end{aligned}
\end{equation}
We remark that the mass and mixing constraints have only a minor impact on the obtained numerical coefficients. Considering purely anarchic matrices $\bm{Y}_q$ of rank~3 and subject to the constraint $|(\bm{Y}_q)_{ij}|\le y_{\rm max}$, the numerical factors in front of $y_{\rm max}$ in the above relations would be 3.0 in the case of $\mbox{Tr}\,\bm{X}_q^2$ and 1.1 in the case of $\kappa_{t,b}$.

\begin{figure}[t!]
\begin{center}
\vspace{2mm}
\makebox{\includegraphics[width=0.85\columnwidth]{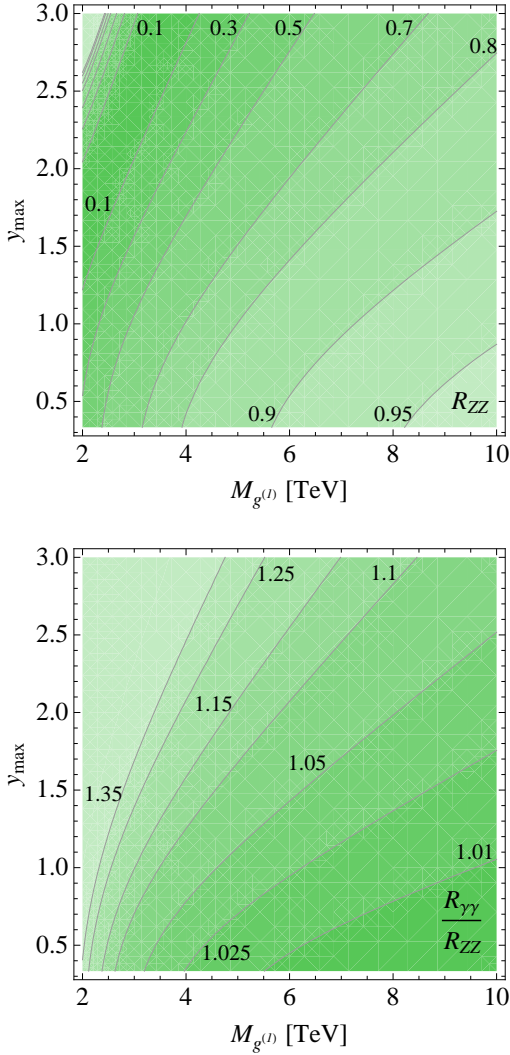}}  
\end{center}
\vspace{-6mm}
\caption{Predictions for $R_{ZZ}$ (upper plot) and $R_{\gamma\gamma}/R_{ZZ}$ (lower plot) in the minimal RS model.}
\label{fig:mRS}
\end{figure}

Our predictions for the ratios $R_{ZZ}$ and $R_{\gamma\gamma}/R_{ZZ}$ in the minimal RS model are shown in Figure~\ref{fig:mRS} as functions of $y_{\rm max}$ and the mass of the lightest KK gluon, $M_{g^{(1)}}\approx 2.45\,M_{\rm KK}$. From the upper plot, we see that $R_{ZZ}$ is strictly smaller than 1 and decreases (increases) with increasing $y_{\rm max}$ ($M_{\rm KK}$). In the region where the ${\cal O}(v_{\rm RS}^2/M_{\rm KK}^2)$ terms are smaller than 1, we find the approximate result 
\begin{equation}\label{RZmRS}
   R_{\rm ZZ}\approx 1 - \frac{v_{\rm RS}^2}{2M_{\rm KK}^2}\,
   \big( 15.6 + 12.9\,y_{\rm max}^2 \big) \,.
\end{equation}
The modifications of the tree-level $ZZh$ coupling (\ref{eq:kappaV}) and of the effective $ggh$ vertex (\ref{kappag}) both suppress $R_{ZZ}$. The deviation of $R_{ZZ}$ from 1 can be particularly large if the Yukawa couplings are allowed to take values near the perturbativity bound $y_{\max}=3$, which was estimated in \cite{Csaki:2008zd} by means of naive dimensional analysis. 
Note that in the upper left corner of the upper plot the corrections to $R_{ZZ}$ become so large that they cancel the SM contribution or even overcompensate it. The lower plot shows that $R_{\gamma\gamma}/R_{ZZ}$ is always larger than 1 and increases with increasing $y_{\rm max}$ and decreasing $M_{\rm KK}$. In other words, RS models predict an enhancement of the $h\to\gamma\gamma$ rate relative to the $h\to ZZ$ rate in comparison with the predictions of the SM. However, the deviation of $R_{\gamma\gamma}/R_{ZZ}$ from 1 is much smaller than for $R_{ZZ}$. Approximately, we obtain
\begin{equation}\label{RgaZmRS}
   \frac{R_{\rm \gamma\gamma}}{R_{\rm ZZ}}
   \approx 1 + \frac{v_{\rm RS}^2}{2M_{\rm KK}^2}\,\big( 0.7 + 4.1\,y_{\rm max}^2 \big) \,,
\end{equation}
which implies that this observable is dominated by the effects of KK fermion loops contributing to $h\to\gamma\gamma$. Combining (\ref{RZmRS}) and (\ref{RgaZmRS}) indicates that the ratio $R_{\gamma\gamma}$ itself is predicted to be smaller than 1 in the minimal RS model, and that its contour plot resembles closely that of $R_{ZZ}$. 

Let us assume that a future measurement of the Higgs production cross section in the $ZZ$ channel yields a value not smaller than 0.8 times the SM cross section. For $y_{\rm max}=3$, which is the value assumed in most phenomenological analyses of flavor-physics constraints in RS models to date, we would then conclude that the mass of the lightest KK gluon must be larger than 11\,TeV. If we impose the stronger bounds $|(\bm{Y}_q)_{ij}|\le 1.5$ or even $\le 1$ on the entries of the Yukawa matrices, we still obtain lower bounds of 6.5\,TeV or 5\,TeV on the KK gluon mass, respectively. These values exceed the masses accessible by direct production at the LHC. It has been estimated that KK gluons with masses up to $(3-4)$\,TeV can be probed with $100\,{\rm fb}^{-1}$ of data collected at $\sqrt{s}=14$\,TeV \cite{Agashe:2006hk,Lillie:2007yh}, while for neutral gauge bosons the reach extends only up to about 2\,TeV \cite{Agashe:2007ki,Agashe:2008jb}. If, on the other hand, a suppression relative to the SM cross section were found in the production of the Higgs boson with subsequent decays into the clean $ZZ$ and $\gamma\gamma$ final states, then based on Figure~\ref{fig:mRS} these measurements could be translated into parameter ranges in the $M_{\rm KK}-y_{\rm max}$ plane. For example, if the one were to observe a reduction of the cross section for Higgs production in the $ZZ$ channel by a factor of 2 (i.e.\ $R_{ZZ}=0.5$), then this would hint at a mass range between 3 and 6\,TeV for the lightest KK gluon, assuming $y_{\rm max}$ lies between 1 and 3.

\begin{figure}[t!]
\begin{center}
\vspace{2mm}
\makebox{\includegraphics[width=0.85\columnwidth]{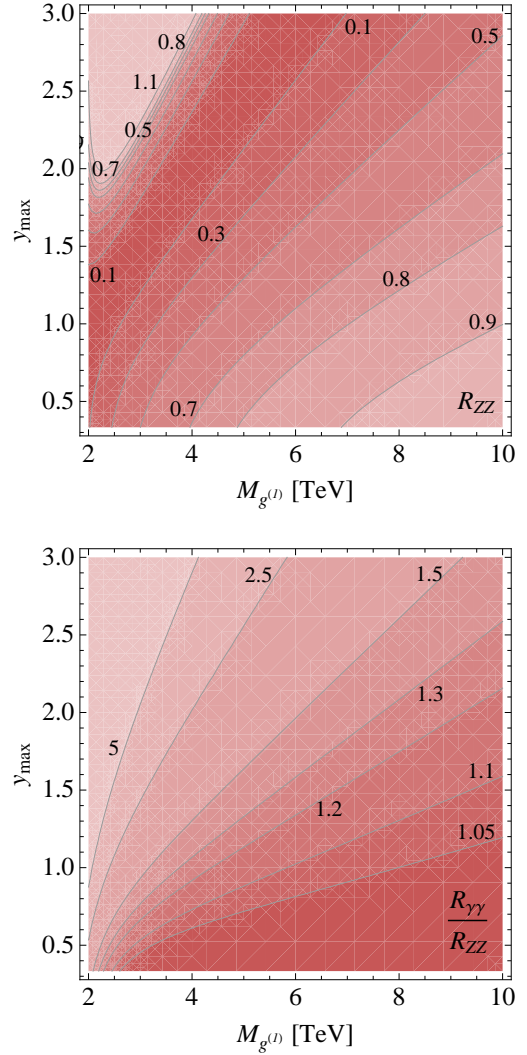}}  
\end{center}
\vspace{-6mm}
\caption{Predictions for $R_{ZZ}$ (upper plot) and $R_{\gamma\gamma}/R_{ZZ}$ (lower plot) in the RS model with custodial protection.}
\label{fig:cRS}
\end{figure}

It is interesting to consider also an extended version of the minimal RS model, in which the bulk gauge group is enlarged to include a custodial symmetry. When the SM gauge symmetry $SU(2)_L\times U(1)_Y$ is replaced by $SU(2)_L\times SU(2)_R\times U(1)_X\times P_{LR}$ in the bulk in order to protect the $T$ parameter \cite{Agashe:2003zs}, we find that the coefficient $(L-1)$ in (\ref{eq:kappav}), (\ref{eq:kappaV}) and (\ref{eq:kappaga}) gets replaced by $2L$, and $\kappa_Z=\kappa_W$ with $m_V=m_W$ in (\ref{eq:kappaV}). A significant reduction of the tree-level contributions to the $Z b_L\bar b_L$ vertex \cite{Agashe:2006at} and its flavor-changing counterparts \cite{Blanke:2008zb} can be accomplished by extending the quark content of the model. In the simplest flavor-anarchic setup considered in \cite{Casagrande:2010si}, this leads to the appearance of 15 up-type quarks, 9 down-type quarks, and 9 exotic quarks with electric charge $5/3$ at each KK level, compared to 6 up-type and 6 down-type quarks in the minimal RS model. The enhanced multiplicity of fermionic degrees of freedom has several important consequences. In (\ref{kappat}) and (\ref{kappab}), the terms that explicitly involve $\bm{Y}_q$ are enhanced by about a factor of 2, and the $\bm{\delta}_{U,u}$ matrices must be replaced by matrices $\bm{\Phi}_{U,u}$ defined in \cite{Casagrande:2010si}. Employing again a large sample of parameter points, we find that the numerical coefficients entering $\kappa_t$ and $\kappa_b$ in (\ref{eq:mRS}) are thereby changed to 3.7 and 2.4, respectively. The higher multiplicity of states has an even larger effect on the expression for $\kappa_g$ in (\ref{kappag}), which can be incorporated by multiplying the term $\sum_{q=u,d} {\rm Tr}\,f(\bm{X}_q)$ by 11/4. Similar modifications apply to $\kappa_\gamma$ in (\ref{eq:kappaga}). A more detailed discussion of the differences between the minimal and the custodial RS model will be presented elsewhere \cite{inprep}.
 
In Figure~\ref{fig:cRS}, we display $R_{ZZ}$ and $R_{\gamma\gamma}/R_{ZZ}$ in the custodial RS model as a function of the lightest KK gluon mass and $y_{\rm max}$. Notice that the qualitative behavior of both quantities parallels that found in the minimal model, but that the deviations of the two ratios from 1 are significantly more pronounced. This feature is a direct consequence of the higher multiplicity of fermionic states in the custodial RS model. In the region where the ${\cal O}(v_{\rm RS}^2/M_{\rm KK}^2)$ terms are smaller than 1, we obtain the approximations
\begin{equation}
\begin{aligned}
   R_{\rm ZZ} 
   &\approx 1 - \frac{v_{\rm RS}^2}{2M_{\rm KK}^2}\,\big( 23.5 + 36.0\,y_{\rm max}^2 \big) \,, \\ 
   \frac{R_{\rm \gamma\gamma}}{R_{\rm ZZ}} 
   &\approx 1 + \frac{v_{\rm RS}^2}{2M_{\rm KK}^2}\,\big( \!-\!7.5 + 23.8\,y_{\rm max}^2 \big) \,.
\end{aligned}
\end{equation}
However, in this case there are also large regions of parameters space where these approximations breaks down. Notice in particular that in the upper left corner of the upper plot the corrections to $R_{ZZ}$ become so large that they overcompensate the SM contribution and $\kappa_g$ becomes negative. In this exceptional region of parameter space, it is possible that the ratio $R_{\gamma\gamma}$ in (\ref{Rgg}) becomes larger than 1, meaning that the production cross section times the $h\to\gamma\gamma$ branching ratio can be larger than in the SM. From the figure it is evident that a measurement of the Higgs production cross section in the $ZZ$ or $\gamma\gamma$ channels close to the SM cross sections would impose severe constraints on the KK mass scale, which would be far stronger than the bounds that could be derived from a non-observation of KK particles at the LHC. In fact, these bounds are so strong that, barring fine-tuning, a non-observation of significant modifications of the cross sections would exclude this version of the custodial RS model over most over the interesting region of its parameter space.

\subsubsection{Conclusions}

We have pointed out that a Higgs-boson discovery at the LHC will lead to stringent constraints on both the structure and parameter space of extensions of the SM with a warped extra dimension. In particular, precision measurements of the individual $pp\to h\to f$ cross sections provide excellent probes of new physics, even if the mass scale of KK excitations lies above $(3-4)$~TeV and thus outside the reach for direct production at the LHC. We have emphasized that RS models predict significant modifications of the effective $ggh$ and $\gamma\gamma h$ couplings due to loop diagrams involving a high multiplicity of virtual KK fermions or $W$ bosons. These effects lead to a reduction of the $gg\to h$ production cross section at the LHC, while the branching fraction for $h\to\gamma\gamma$ is enhanced. The branching ratios for decays into $ZZ$, $WW$, and  $b\bar b$ pairs receive only small corrections. They are dominated by the tree-level couplings of the Higgs boson to fermions and gauge bosons, which are SM-like. As a result, the cross section times branching ratios for $pp\to h\to\gamma\gamma$ and $pp\to h\to ZZ, WW$ are predicted to be reduced, but the suppression is more marked in the latter case. This pattern resembles the preliminary results of the ATLAS and CMS collaborations, which indicate that a deviation from the SM predictions should be of modest size only. Once the Higgs boson has been discovered and its branching ratios to di-boson final states have been measured with some precision, these results will have profound implications for RS models, which will either provide further support to the idea of extra dimensions even in cases where no KK particles are discovered, or else push the mass scale for KK excitations into the 10\,TeV range.

\end{document}